%% file: main.tex
\documentclass{vgtc}                     

\ifpdf
  \pdfoutput=1\relax                   
  \usepackage{graphicx}                
  \DeclareGraphicsExtensions{.pdf,.png,.jpg,.jpeg} 
\else
  \usepackage{graphicx}                
  \DeclareGraphicsExtensions{.eps}     
\fi%

\onlineid{1027}

\vgtccategory{Research}

\vgtcpapertype{theory/model}

\title{What Exactly is an Insight? A Literature Review}

\author{%
  Leilani Battle\thanks{e-mail: leibatt@cs.washington.edu}\\ %
        \scriptsize University of Washington
  \and Alvitta Ottley\thanks{e-mail: alvitta@wustl.edu}\\ %
        \scriptsize Washington University in St. Louis
}

\authorfooter{
  \item Leilani Battle is with the University of Washington, Seattle.
        E-mail: leibatt@uw.edu.
  \item Alvitta Ottley is with Washington University in St. Louis.
        E-mail: alvitta@wustl.edu.
}

\abstract{%
  \input{sections/abstract.tex}
}


\graphicspath{{figs/}{figures/}{pictures/}{images/}{./}} 

\usepackage{mathptmx}                  

\usepackage{xspace}

\makeatletter
\DeclareRobustCommand\onedot{\futurelet\@let@token\@onedot}
\def\@onedot{\ifx\@let@token.\else.\null\fi\xspace}

\makeatother

\usepackage[table,xcdraw,svgnames]{xcolor}
\usepackage{hyperref}
\usepackage{enumitem}
\usepackage{multirow}
\usepackage{comment}
\usepackage{changepage}
\usepackage{subcaption}

\newif\ifnotes
\notesfalse

\newcommand{\revised}[1]{\ifnotes{\leavevmode\color{blue} #1}\else{#1}\fi}

\begin{document}


\firstsection{Introduction}

\maketitle

\input{sections/introduction}
\input{sections/background}
\input{sections/discussion}
\input{sections/conclusion}

	

\acknowledgments{\revised{This work was supported in part by the NSF through award numbers IIS-2141506, IIS-2142977, and OAC-2118201.}}

\bibliographystyle{abbrv-doi-hyperref}

\bibliography{vis-task-theory2}



\end{document}

%% file: sections/abstract.tex
Insights are often considered the ideal outcome of visual analysis sessions. However, there is no single definition of what an insight is. Some scholars define insights as correlations, while others define them as hypotheses or aha moments. This lack of a clear definition can make it difficult to build visualization tools that effectively support insight discovery. In this paper, we contribute a comprehensive literature review that maps the landscape of existing insight definitions. We summarize key themes regarding how insight is defined, with the goal of helping readers identify which definitions of insight align closely with their research and tool development goals. Based on our review, we also suggest interesting research directions, such as synthesizing a unified formalism for insight and connecting theories of insight to other critical concepts in visualization research.

%% file: sections/introduction.tex
\label{sec:introduction}
\vspace{-1mm}

Card, Mackinlay, and Shneiderman assert that the goal of information visualization is to amplify the user’s cognition~\cite{card1999using}. 
To this end, many visualization tools aim to amplify cognition by helping users extract reliable \emph{insights} from their data~\cite{north_comparison_2011,chang_defining_2009} (e.g., to increase insight generation rates~\cite{liu_effects_2014,zgraggen2017progressive} or to reduce the incidence of false discoveries~\cite{zgraggen_investigating_2018,zhao2017controlling}).
However, to enhance \emph{insight discovery},  it is crucial to have a clear understanding of what insights are and how they can be discovered.
Yet, the literature presents different definitions of ``insights''~\cite{kandogan2016grounded}.
For example, Sariaya et al. define insight as a \emph{unit of discovery}~\cite{saraiya_insight-based_2005} while Gomez et al.~\cite{gomez_insight-_2014} and Guo et al.~\cite{guo_case_2016} argue that insights are \emph{hypotheses}. Demiralp et al.~\cite{demiralp_foresight_2017} view insights as \emph{statistical data properties} such as correlations between variables. Chen et al. suggest that insights are \emph{links between statistical data properties and user domain knowledge}~\cite{yang_chen_toward_2009}. Additionally, Chang et al.~\cite{chang_defining_2009} draw on psychology and cognitive science to define insights as \emph{``aha'' moments} as well as \emph{links between units of knowledge}.

Defining insights is a critical challenge for the visualization community because it can significantly influence tool development trajectories. Depending on the chosen definition of insights, developers might pursue vastly different approaches. 
For example, if insights are hypotheses, developers might design visualization tools to maximize the generation of verifiable hypotheses such as significant data correlations or outliers (e.g., \cite{demiralp_foresight_2017}). In contrast, if insights are ``aha'' moments, a researcher may instead design creativity support tools~\cite{shneiderman2007creativity} or problem-solving tools~\cite{cybulski2015digital} to maximize user creativity and inspiration. Alternatively, if insights are links to users' domain knowledge, researchers may strive to build knowledge management tools for tracking what users have learned over time (e.g., \cite{gotz_interactive_2006}).
This raises the question of which insight definition is the ``right'' one to use when developing a new insight-based study or visualization tool.
\revised{Further, what factors should influence this choice?}

\revised{This paper summarizes the key considerations from the visualization literature regarding how insights are defined and how to \emph{design applications} around these definitions.}
Our aim is to help researchers and developers choose (\revised{or create}) the definitions that best suit their \revised{design} objectives. Additionally, we identify areas of interest for future research, including the opportunity to contribute new insight definitions, integrate existing definitions, or establish connections between insight definitions and related concepts in the literature.

In summary, this paper makes the following contributions:
\begin{itemize}[nosep]
\item We present a \textbf{literature review on the definition of insights}.
\item We \textbf{categorize the definitions and their pros/cons} to help readers identify suitable definitions to motivate their work.
\item We \textbf{present interesting research directions} in insight- and theory-based visualization research motivated by this work.
\end{itemize}

%% file: sections/background.tex
\section{How are Insights Defined in the Literature?}
\label{sec:background}
\vspace{-1mm}

\subsection{Overview}

\begin{table*}[]
\centering
{\small
\begin{tabular}{lcccccc}
\hline
\textbf{Categorizations} & \multicolumn{3}{c}{\textit{\textbf{Knowledge Building vs. "Eureka" Moments}}}    & \multicolumn{3}{c}{\textit{\textbf{Insight Sources}}}                                                                                                                                                     \\ \hline
                                 & \multicolumn{3}{c}{\cite{chang_defining_2009,plaisant_promoting_2008,pousman_casual_2007,shneiderman_inventing_2002}}                                                           & \multicolumn{3}{c}{\cite{choe_characterizing_2015,gomez_insight-_2014,liu_effects_2014,north_toward_2006,pousman_casual_2007,saraiya_evaluation_2004, saraiya_insight-based_2005,smuc_score_2009,yi_understanding_2008,zgraggen_investigating_2018}}                                                                                                                                                                                    \\ \hline
\textbf{Definitions}     & \textit{\textbf{Utterances}}  & \multicolumn{2}{c}{\textit{\textbf{Data Facts}}} & \textit{\textbf{Hypotheses}} & \multicolumn{2}{c}{\textit{\textbf{Knowledge Links}}}                                                                                                                      \\ \hline
\multirow{2}{*}{}                & \multirow{2}{*}{\cite{north_comparison_2011,plaisant_promoting_2008,saraiya_evaluation_2004,saraiya_insight-based_2005,saraiya_insight-based_2006,zgraggen2017progressive,zgraggen_investigating_2018}}           & \textit{General}    & \textit{Recommendations}   & \multirow{2}{*}{\cite{gomez_insight-_2014,liu_effects_2014}}          & \textit{General}                                                                  & \textit{Knowledge Graphs}                                                              \\
                                 &                               & \cite{yang_chen_toward_2009}                & \cite{cui_datasite_2019,demiralp_foresight_2017,srinivasan_augmenting_2019,vartak_seedb_2015,zeng2022evaluation}                       &                              & \cite{amar_knowledge_2005,choe_characterizing_2015,gotz_characterizing_2009,gotz_interactive_2006,green_visual_2008,north_toward_2006,pousman_casual_2007,sacha_knowledge_2014,yi_understanding_2008}                                                                              & \cite{he_characterizing_2020,kandogan2018towards,mathisen_insideinsights_2019,rind_task_2016,shrinivasan_connecting_2009,smuc_score_2009,shrinivasan_supporting_2008,willett2011commentspace}                                                                                  \\ \hline
\textbf{Scope}           & \multicolumn{2}{c}{\textit{\textbf{Taxonomies \& Typologies}}}    & \multicolumn{2}{c}{\textit{\textbf{Frameworks}}}          & \textit{\textbf{\begin{tabular}[c]{@{}c@{}}Processes \&\\ Patterns\end{tabular}}} & \textit{\textbf{\begin{tabular}[c]{@{}c@{}}Declarative\\ Specifications\end{tabular}}} \\ \hline
                                 & \multicolumn{2}{c}{\cite{amar_low-level_2005,brehmer_multi-level_2013,gathani2022grammar, kang_examining_2012,zgraggen_investigating_2018}}                             & \multicolumn{2}{c}{\cite{gotz_characterizing_2009,lam_bridging_2018}}                                    & \cite{battle_characterizing_2019,guo_case_2016,yi_understanding_2008}                                                                                & \cite{andrienko2006exploratory,bertin1983semiology,codd1970relational}                                                                                     \\ \hline
\textbf{\begin{tabular}[c]{@{}l@{}}Potential\\ Applications\end{tabular}}    & \multicolumn{1}{l}{\textit{\textbf{Life Logging}}} & \multicolumn{1}{l}{\textit{\textbf{\begin{tabular}[c]{@{}l@{}}Bioinformatics/\\ Medicine\end{tabular}}}} & \multicolumn{1}{l}{\textit{\textbf{\begin{tabular}[c]{@{}l@{}}Problem Solving/\\ Creativity Support\end{tabular}}}} & \multicolumn{1}{l}{\textit{\textbf{Provenance}}} & \multicolumn{1}{l}{\textit{\textbf{\begin{tabular}[c]{@{}l@{}}False Discoveries/\\ Multiple Comparisons\end{tabular}}}} & \multicolumn{1}{l}{\textit{\textbf{Grammars}}}                                         \\ \hline
                                 & \multicolumn{1}{l}{\cite{choe_characterizing_2015}}                             & \multicolumn{1}{l}{\cite{he_characterizing_2020,saraiya_insight-based_2005}}                                                                                  & \multicolumn{1}{l}{\cite{cybulski2015digital,shneiderman2007creativity}}                                                                                            & \multicolumn{1}{l}{\cite{xu_survey_2020,battle_characterizing_2019}}                         & \multicolumn{1}{l}{\cite{zgraggen_investigating_2018,zhao2017controlling}}                                                                                               & \multicolumn{1}{l}{\cite{gathani2022grammar,kandogan2018towards,satyanarayan2017vegalite,wickham2011ggplot2,wilkinson2012grammar}}                                                               \\ \hline

\end{tabular}
}
\caption{\revised{We survey three approaches to analyzing insights in the literature: their categorizations, definitions, and scope within visual analysis tasks. We also discuss potential applications of existing insight methodologies. References are organized by relevance to analysis approach.}}
\label{tab:overview}
\end{table*}

Existing research places a strong emphasis on \emph{insight discovery} during visual analysis and exploration tasks~\cite{north_toward_2006,chang_defining_2009,guo_case_2016}
For example, researchers often test visual analysis tools by the quantity, accuracy, and quality of insights users generate while using them~\cite{saraiya_insight-based_2005,north_comparison_2011,liu_effects_2014,battle_characterizing_2019,zgraggen_investigating_2018}.
In this section, we summarize existing definitions and characteristics of insight and highlight recurring overlaps and themes.

\vspace{-1mm}
\paragraph{Review Process.} We conducted keyword searches for ``visualization task'' and ``visualization insight'' in Google Scholar. We also reviewed the proceedings of VIS and EuroVis from 2013 to 2023, including papers describing visualization objectives, tasks, and provenance, since they are often discussed alongside insights.
This generated an initial list of 125 papers. We reviewed each paper to verify its relevance to visualization and whether it focused on defining, analyzing, or supporting insight discovery, e.g., ``But what, exactly, is insight? How can it be measured and evaluated?''~\cite{north_comparison_2011}. For each relevant paper, we reviewed its list of references to identify papers we may have missed. These steps yielded a list of 38 papers. With feedback from colleagues/reviewers, we extended it to include their suggestions, producing a final list of 41 papers.
We analyzed how insight was defined in each paper, focusing on high-level themes and key characteristics of insights. Additionally, we cite synergistic ideas when relevant, e.g., tasks and visualization recommendations.

\vspace{-1mm}
\paragraph{\revised{Review Structure.}} \revised{We provide a brief overview of the literature surveyed in \autoref{tab:overview}. Our literature review reveals three major themes in how insights are analyzed:  \emph{categorizing}, \emph{defining}, and \emph{scoping} insights during visual analysis.  In the remainder of this section, we summarize each analysis method and how the resulting contributions may influence the design of relevant, insight-driven applications. Potential example applications are also described in the text and cited in \autoref{tab:overview}.}

\vspace{-1mm}
\subsection{Categories of Insight.}
\label{sec:background:insight:categories}
\vspace{-1mm}

The prior work details several high-level categories of insights \revised{that visualization tools can support}.
The first set of categorizations we observe distinguishes between instantaneous sparks and long-term knowledge building. Chang et al.~\cite{chang_defining_2009} distinguish between a ``knowledge-building insight,'' or information directly extending a user's existing knowledge structures, and ``spontaneous insight,'' or a ``eureka'' moment that 
connects loosely related knowledge structures. This distinction is similar to ``directed'' versus ``unexpected'' insights as proposed by Saraiya et al.~\cite{saraiya_insight-based_2005}. Chang et al.~\cite{chang_defining_2009} argue that knowledge-building is typically the focus of visualization and visual analytics work~\cite{chang_defining_2009}, though some argue the opposite~\cite{plaisant_promoting_2008,pousman_casual_2007,shneiderman_inventing_2002}.

Alternatively, categorizations \revised{may} focus on the \emph{source} of insights, such as the input dataset, social structures of the analyst, or an analyst's external domain knowledge.
We summarize these categorizations in \autoref{tab:background:categories}, \revised{where overlaps in categories share the same row(s) of the table}.
Saraiya et al. define four categories of data-driven insights~\cite{saraiya_insight-based_2005,saraiya_evaluation_2004}: overall distributions, patterns, grouping, and detail.
Choe et al.~\cite{choe_characterizing_2015} extend these ideas by providing more granular categorizations of data-driven insights, such as distinguishing distributions versus data summaries or correlations versus trends in identifying patterns.
Zgraggen et al.~\cite{zgraggen_investigating_2018} follow a similar structure but focus on categorizing the different ways in which people make data comparisons to extract data-driven insights.
Guo et al.~\cite{guo_case_2016} and Gomez et al.~\cite{gomez_insight-_2014} abstract these ideas into higher-level categories that connect the data facts emphasized by Choe et al.~\cite{choe_characterizing_2015} with more generalization-focused insights observed by Zgraggen et al.~\cite{zgraggen_investigating_2018} and a separate category of data-driven insights known as \emph{hypotheses}, i.e., conjectures about the dataset that can be tested through subsequent confirmatory analyses.
Liu and Heer propose seven insight categories for analyzing exploratory visual analysis outcomes~\cite{liu_effects_2014}: ``observation,'' ``generalization,'' ``hypothesis,'' ``question,'' ``recall,'' ``interface,'' and ``simulation,'' where we find that the simulation category was not proposed in prior insight studies. The Liu and Heer taxonomy~\cite{liu_effects_2014} acts as a superset of sorts and overlaps significantly with categories proposed in the literature, such as those by Saraiya et al.~\cite{saraiya_evaluation_2004,saraiya_insight-based_2005,saraiya_insight-based_2006} and North~\cite{north_comparison_2011}, Smuc et al.~\cite{smuc_score_2009}, Gomez et al.~\cite{gomez_insight-_2014} and Guo et al.~\cite{guo_case_2016}, and Yi et al. ~\cite{yi_understanding_2008}.  Note that these categories are not mutually exclusive and have been found to co-occur~\cite{smuc_score_2009}.

However, data-driven insights are not the only insights an analyst may uncover. For example, Gotz et al.~\cite{gotz_interactive_2006}, Pousman et al.~\cite{pousman_casual_2007}, Liu and Heer~\cite{liu_effects_2014}, and Choe et al.~\cite{choe_characterizing_2015} observe that analysts often connect what they see in the data with their own knowledge and experiences, i.e., with \emph{domain knowledge} that exists outside the target dataset. Pousman et al. broaden this view to support other kinds of insights that may not be purely data-driven, in particular ``awareness insight,'' ``social insight,'' and ``reflective insight''~\cite{pousman_casual_2007}. Furthermore, Smuc et al.~\cite{smuc_score_2009} and Liu and Heer~\cite{liu_effects_2014} observe that users may also gain insights into how to improve the visual analysis tool they are interacting with, yielding UI-driven insights.

\begin{table*}[]
\centering
{\small
\begin{tabular}{lllllll}
\hline
\textbf{Motivation}                                                           & \textbf{Pousman et al.~\cite{pousman_casual_2007}}                                                                & \textbf{Liu \& Heer~\cite{liu_effects_2014}}                                                                  & \textbf{Guo et al.~\cite{guo_case_2016}}                                                                  & \textbf{Choe et al.~\cite{choe_characterizing_2015}}                                                             & \textbf{Saraiya et al.~\cite{saraiya_insight-based_2005}}                                                        & \textbf{Zgraggen et al.~\cite{zgraggen_investigating_2018}}                                        \\ \hline
\multicolumn{1}{|l|}{\cellcolor[HTML]{CFE2F3}}                                & \multicolumn{1}{l|}{\cellcolor[HTML]{CFE2F3}}                                   & \multicolumn{1}{l|}{\cellcolor[HTML]{CFE2F3}}                                 & \multicolumn{1}{l|}{\cellcolor[HTML]{CFE2F3}}                                 & \multicolumn{1}{l|}{\cellcolor[HTML]{CFE2F3}Detail}                       & \multicolumn{1}{l|}{\cellcolor[HTML]{CFE2F3}Detail}                     &                                                          \\ \cline{5-6}
\multicolumn{1}{|l|}{\cellcolor[HTML]{CFE2F3}}                                & \multicolumn{1}{l|}{\cellcolor[HTML]{CFE2F3}}                                   & \multicolumn{1}{l|}{\cellcolor[HTML]{CFE2F3}}                                 & \multicolumn{1}{l|}{\cellcolor[HTML]{CFE2F3}}                                 & \multicolumn{1}{l|}{\cellcolor[HTML]{CFE2F3}Trend}                        &                                                                         &                                                          \\ \cline{5-7} 
\rowcolor[HTML]{CFE2F3} 
\multicolumn{1}{|l|}{\cellcolor[HTML]{CFE2F3}}                                & \multicolumn{1}{l|}{\cellcolor[HTML]{CFE2F3}}                                   & \multicolumn{1}{l|}{\cellcolor[HTML]{CFE2F3}}                                 & \multicolumn{1}{l|}{\cellcolor[HTML]{CFE2F3}}                                 & \multicolumn{1}{l|}{\cellcolor[HTML]{CFE2F3}Correlation}                  & \multicolumn{1}{l|}{\cellcolor[HTML]{CFE2F3}Patterns}                   & \multicolumn{1}{l|}{\cellcolor[HTML]{CFE2F3}Correlation} \\ \cline{5-7} 
\multicolumn{1}{|l|}{\cellcolor[HTML]{CFE2F3}}                                & \multicolumn{1}{l|}{\cellcolor[HTML]{CFE2F3}}                                   & \multicolumn{1}{l|}{\cellcolor[HTML]{CFE2F3}}                                 & \multicolumn{1}{l|}{\cellcolor[HTML]{CFE2F3}}                                 & \multicolumn{1}{l|}{\cellcolor[HTML]{CFE2F3}Data Summary}                 &                                                                         &                                                          \\ \cline{5-7} 
\rowcolor[HTML]{CFE2F3} 
\multicolumn{1}{|l|}{\cellcolor[HTML]{CFE2F3}}                                & \multicolumn{1}{l|}{\cellcolor[HTML]{CFE2F3}}                                   & \multicolumn{1}{l|}{\cellcolor[HTML]{CFE2F3}}                                 & \multicolumn{1}{l|}{\cellcolor[HTML]{CFE2F3}}                                 & \multicolumn{1}{l|}{\cellcolor[HTML]{CFE2F3}Distribution}                 & \multicolumn{1}{l|}{\cellcolor[HTML]{CFE2F3}Overall Distributions}      & \multicolumn{1}{l|}{\cellcolor[HTML]{CFE2F3}Shape}       \\ \cline{5-7} 
\multicolumn{1}{|l|}{\cellcolor[HTML]{CFE2F3}}                                & \multicolumn{1}{l|}{\cellcolor[HTML]{CFE2F3}}                                   & \multicolumn{1}{l|}{\multirow{-6}{*}{\cellcolor[HTML]{CFE2F3}Observation}}    & \multicolumn{1}{l|}{\multirow{-6}{*}{\cellcolor[HTML]{CFE2F3}Fact}}           & \multicolumn{1}{l|}{\cellcolor[HTML]{CFE2F3}Outlier}                      &                                                                         &                                                          \\ \cline{3-7} 
\rowcolor[HTML]{CFE2F3} 
\multicolumn{1}{|l|}{\cellcolor[HTML]{CFE2F3}}                                & \multicolumn{1}{l|}{\cellcolor[HTML]{CFE2F3}}                                   & \multicolumn{1}{l|}{\cellcolor[HTML]{CFE2F3}}                                 & \multicolumn{1}{l|}{\cellcolor[HTML]{CFE2F3}}                                 & \multicolumn{1}{l|}{\cellcolor[HTML]{CFE2F3}}                             & \multicolumn{1}{l|}{\cellcolor[HTML]{CFE2F3}}                           & \multicolumn{1}{l|}{\cellcolor[HTML]{CFE2F3}Mean}        \\ \cline{7-7} 
\rowcolor[HTML]{CFE2F3} 
\multicolumn{1}{|l|}{\cellcolor[HTML]{CFE2F3}}                                & \multicolumn{1}{l|}{\cellcolor[HTML]{CFE2F3}}                                   & \multicolumn{1}{l|}{\cellcolor[HTML]{CFE2F3}}                                 & \multicolumn{1}{l|}{\cellcolor[HTML]{CFE2F3}}                                 & \multicolumn{1}{l|}{\cellcolor[HTML]{CFE2F3}}                             & \multicolumn{1}{l|}{\cellcolor[HTML]{CFE2F3}}                           & \multicolumn{1}{l|}{\cellcolor[HTML]{CFE2F3}Variance}    \\ \cline{7-7} 
\rowcolor[HTML]{CFE2F3} 
\multicolumn{1}{|l|}{\cellcolor[HTML]{CFE2F3}}                                & \multicolumn{1}{l|}{\multirow{-9}{*}{\cellcolor[HTML]{CFE2F3}Analytic insight}} & \multicolumn{1}{l|}{\multirow{-3}{*}{\cellcolor[HTML]{CFE2F3}Generalization}} & \multicolumn{1}{l|}{\multirow{-3}{*}{\cellcolor[HTML]{CFE2F3}Generalization}} & \multicolumn{1}{l|}{\multirow{-3}{*}{\cellcolor[HTML]{CFE2F3}Comparison}} & \multicolumn{1}{l|}{\multirow{-3}{*}{\cellcolor[HTML]{CFE2F3}Grouping}} & \multicolumn{1}{l|}{\cellcolor[HTML]{CFE2F3}Ranking}     \\ \cline{2-7} 
\multicolumn{1}{|l|}{\cellcolor[HTML]{CFE2F3}}                                & \multicolumn{1}{l|}{}                                                           & \multicolumn{1}{l|}{\cellcolor[HTML]{CFE2F3}Hypothesis}                       & \multicolumn{1}{l|}{\cellcolor[HTML]{CFE2F3}Hypothesis}                       &                                                                           &                                                                         &                                                          \\ \cline{3-4}
\multicolumn{1}{|l|}{\cellcolor[HTML]{CFE2F3}}                                & \multicolumn{1}{l|}{}                                                           & \multicolumn{1}{l|}{\cellcolor[HTML]{CFE2F3}Question}                         &                                                                               &                                                                           &                                                                         &                                                          \\ \cline{3-3}
\multicolumn{1}{|l|}{\multirow{-12}{*}{\cellcolor[HTML]{CFE2F3}Data-driven}}  & \multicolumn{1}{l|}{}                                                           & \multicolumn{1}{l|}{\cellcolor[HTML]{CFE2F3}Simulation}                       &                                                                               &                                                                           &                                                                         &                                                          \\ \cline{1-3} \cline{5-5}
\multicolumn{1}{|l|}{\cellcolor[HTML]{FCE5CD}}                                & \multicolumn{1}{l|}{\cellcolor[HTML]{FCE5CD}Reflective insight}                 & \multicolumn{1}{l|}{\cellcolor[HTML]{FCE5CD}Recall}                           & \multicolumn{1}{l|}{}                                                         & \multicolumn{1}{l|}{\cellcolor[HTML]{FCE5CD}Self-reflection}              &                                                                         &                                                          \\ \cline{2-3} \cline{5-5}
\multicolumn{1}{|l|}{\multirow{-2}{*}{\cellcolor[HTML]{FCE5CD}Domain-driven}} & \multicolumn{1}{l|}{\cellcolor[HTML]{FCE5CD}Awareness insight}                  &                                                                               &                                                                               &                                                                           &                                                                         &                                                          \\ \cline{1-2}
\multicolumn{1}{|l|}{\cellcolor[HTML]{EAD1DC}Socially-driven}                 & \multicolumn{1}{l|}{\cellcolor[HTML]{EAD1DC}Social insight}                     &                                                                               &                                                                               &                                                                           &                                                                         &                                                          \\ \cline{1-3}
\multicolumn{1}{|l|}{\cellcolor[HTML]{EAD1DC}UI-driven}                       & \multicolumn{1}{l|}{}                                                           & \multicolumn{1}{l|}{\cellcolor[HTML]{EAD1DC}Interface}                        &                                                                               &                                                                           &                                                                         &                                                          \\ \hline
Cited By                                                                       & \cite{brehmer_multi-level_2013,yi_understanding_2008}                                                                         & \cite{gathani2022grammar,liu_effects_2014,zgraggen_investigating_2018,zgraggen2017progressive}                                                                           & \begin{tabular}[c]{@{}l@{}}\cite{he_characterizing_2020,gathani2022grammar,xu_survey_2020,battle_characterizing_2019,kandogan2018towards}\\ \cite{zgraggen_investigating_2018,zgraggen2017progressive,zhao2017controlling}\end{tabular}                                                                          & \cite{he_characterizing_2020,xu_survey_2020}                                                                      & \begin{tabular}[c]{@{}l@{}}\cite{he_characterizing_2020,battle_characterizing_2019,kandogan2018towards,zgraggen_investigating_2018,rind_task_2016,choe_characterizing_2015}\\ \cite{sacha_knowledge_2014, guo_case_2016, saraiya_insight-based_2005,north_comparison_2011, chang_defining_2009}\\ \cite{smuc_score_2009,yi_understanding_2008,pousman_casual_2007,saraiya_insight-based_2006,north_toward_2006}\end{tabular}                                                                & \cite{battle_characterizing_2019,kandogan2018towards}                                                 \\ \hline
\end{tabular}
}
\caption{We observe significant overlaps in how insights are categorized in the literature. Observed categories are colored and labeled by the source of the insight: the data, external domain knowledge, social structure, or the visualization interface itself. \revised{Overlapping categories share the same row. Papers that cite each categorization are recorded in the bottom row.}\vspace{-5mm}}
\label{tab:background:categories}
\end{table*}

\vspace{-1mm}
\subsubsection{\revised{Using Insight Categorizations to Inform Tool Design}}
\vspace{-1mm}

Each categorization emphasizes different sources of insight that can impact how we design visualization tools. For example, Guo et al.~\cite{guo_case_2016}, Saraiya et al.~\cite{saraiya_insight-based_2005} and Zgraggen et al.~\cite{zgraggen_investigating_2018} emphasize data-driven sources of insight, which may be appropriate for tools designed for rigorous testing and management of key data statistics, e.g., to prevent false discoveries~\cite{zhao2017controlling}. Pousman et al.~\cite{pousman_casual_2007}, Liu and Heer~\cite{liu_effects_2014} and Choe et al.~\cite{choe_characterizing_2015} broaden this view to consider the context of the person performing the analysis, which can be critical for tools used for a domain-specific purpose (e.g., medicine~\cite{saraiya_insight-based_2005, he_characterizing_2020}) or a personal one (e.g., life logging~\cite{choe_characterizing_2015}). With an integrative view of these categorizations, we hope to empower readers to choose a categorization that best suits their research and development needs.

\vspace{-1mm}
\subsection{The Varying Definitions of Insight.}
\label{sec:background:insight:definitions}
\vspace{-1mm}

Although insights are often categorized in similar ways, the literature present inconsistent \emph{definitions} for what constitutes an insight. Are they utterances, statistical correlations, or something more complex? In this section, we summarize the definitions proposed in the literature and discuss the pros and cons that may affect their use.

\vspace{-1mm}
\subsubsection{Insights are Utterances}
\vspace{-1mm}

Some definitions assert that \textbf{insights are utterances}. Saraiya et al. define insight as ``an individual observation about the data by the participant, a unit of discovery,''~\cite{saraiya_insight-based_2005,saraiya_evaluation_2004} which can include ``any data observation that the user mentions'' during in-person lab studies~\cite{saraiya_insight-based_2005,liu_effects_2014,zgraggen_investigating_2018,zgraggen2017progressive}, as well as self-reported insight diaries collected through field studies~\cite{saraiya_insight-based_2006} and competition submissions~\cite{plaisant_promoting_2008}. 
Gomez et al. observe that users may only report a subset of their insights relevant to the study at hand~\cite{gomez_insight-_2014}.
Zgraggen et al. posit that insights may not only be explicitly defined through direct user reporting but also implicitly defined through observation, such as when the user is observed performing an analysis but does not officially report the outcome of this analysis to experimenters~\cite{zgraggen_investigating_2018}. 

\vspace{-1mm}
\paragraph{Pros and Cons.} This definition adopts a stream-of-consciousness view of insights that require experimenters to bear witness to the utterance in order to capture the corresponding insight. On the one hand, this definition is easy to apply in insight-based studies since all that is required is an experimenter to observe a user's utterances. On the other hand, this definition places a significant burden on the experimenter to manually identify and validate utterances, ignoring the potential role that visualization provenance and automation can play in helping to detect insights~\cite{xu_survey_2020}.

\vspace{-1mm}
\subsubsection{Insights are Data Facts}
\vspace{-1mm}

Several works categorize insights in terms of how their calculation supports user hypotheses, claims, and reflections, pointing to a third definition -- \textbf{insights are data facts}.
As shown in \autoref{tab:background:categories}, Choe et al. propose eight insight classes, where six classes are statistical in nature (``trend,'' ``correlation,'' ``data summary,'' ``distribution,'' ``outlier'' and ``comparison'') and two are adapted from existing taxonomies ( ``detail''~\cite{saraiya_insight-based_2005,saraiya_insight-based_2006} and ``self-reflection''~\cite{pousman_casual_2007}).
Zgraggen et al. propose five insight classes, all of which are statistical in nature~\cite{zgraggen_investigating_2018}.
We observe that these statistical insight classes extend those initially proposed by Saraiya et al.~\cite{saraiya_insight-based_2005,saraiya_insight-based_2006}.
Liu and Heer~\cite{liu_effects_2014}, Pousman et al.~\cite{pousman_casual_2007}, and Chen et al.~\cite{yang_chen_toward_2009} group these different statistical representations into a single high-level category, i.e., ``observation,'' ``analytic insight'' and ``data facts,'' respectively. These overlaps suggest that collectively, \emph{data facts} may be a core building block of insights. Chen et al. formalize the relationship between data facts and insights through their Fact Management Framework~\cite{yang_chen_toward_2009}, which provides a theoretical base from which to formalize insights. Building on these ideas, visualization recommendation systems such as ForeSight~\cite{demiralp_foresight_2017}, DataSite~\cite{cui_datasite_2019}, SeeDB~\cite{vartak_seedb_2015}, and Voder~\cite{srinivasan_augmenting_2019} extract statistical patterns and anomalies to strengthen the user's understanding of the data and hopefully guide the user toward new insights.

\vspace{-1mm}
\paragraph{Pros and Cons.} The allure of data facts lies in how easy they are to compute. For example, when insight is defined as a linear correlation, it becomes straightforward for a visualization recommendation system to automatically recommend data-driven insights by testing every pair of variables for a correlation~\cite{zeng2022evaluation}. Furthermore, a statistical or mathematical representation of insight enables researchers to test their \emph{validity}. For example, Zgraggen et al. tested the accuracy of reported insights by mapping each utterance from their user study to corresponding dataset statistics such as mean, variance, and linear correlation~\cite{zgraggen_investigating_2018}.
However, this definition of insight ignores the role of domain knowledge in contextualizing statistical results. For example, a correlation only becomes meaningful when it represents a relationship that matters in the real world, such as a correlation between racial discrimination and incidence of crime~\cite{mathisen_insideinsights_2019} or precipitation and incidence of wildlife strikes~\cite{battle_characterizing_2019}. Thus, we should be cautious when adopting this definition since a naive application could yield spurious and ungrounded results.

\begin{figure}
    \centering
    \includegraphics[width=0.9\columnwidth]{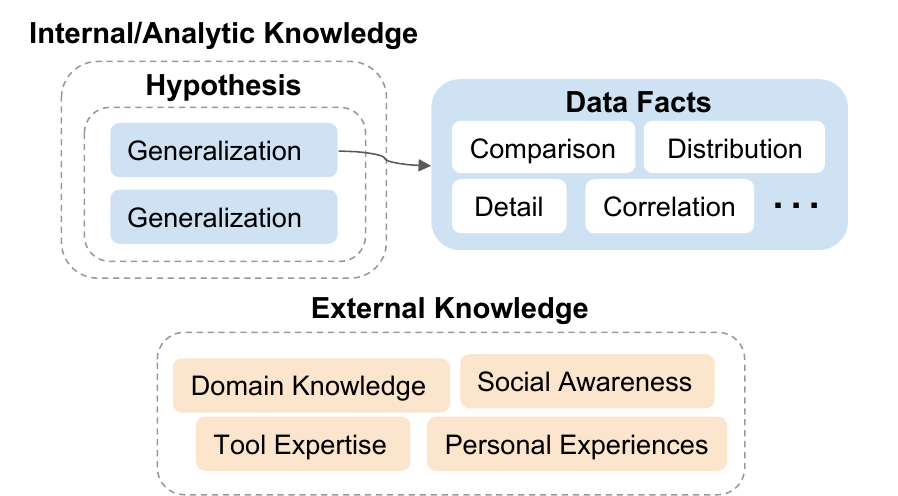}
    \caption{\revised{Insights seem to capture \emph{internal knowledge} extracted from data such as data facts, generalizations of these facts, and hypotheses to be tested. Insights also link internal and \emph{external knowledge} such as domain knowledge, personal experiences, and tool expertise.}\vspace{-5mm}}
    \label{fig:background:definitions:knowledge-units}
\end{figure}

\vspace{-1mm}
\subsubsection{Insights are Hypotheses}
\vspace{-1mm}

The prior work also suggests that \textbf{insights are hypotheses and/or evidence.} For example, to evaluate how study participants perform during open-ended exploration tasks, Gomez et al. label each observed insight from their study as a ``claim,'' i.e., ``a general hypothesis, question, or remark about the data model that is potentially synthesized from multiple observations,'' or as ``evidence,''  such as an observation comprised of ``specific references to data points'' supporting the claim~\cite{gomez_insight-_2014}.
Guo et al. augment this claim-evidence structure to encompass ``facts,'' ``generalizations,'' and ``hypotheses''~\cite{guo_case_2016}, where facts are units of truth about specific entities in the data, generalizations are inferred relationships between observed entities, and hypotheses are claims that facts and generalizations can support.
 Liu and Heer adopt a similar strategy where they observe that analysts' data observations and generalizations can lead to new hypotheses, i.e., insights~\cite{liu_effects_2014}.
Similarly, Sacha et al. observe that ``Analysts try to find evidence that supports or contradicts hypotheses in order to gain knowledge from data''~\cite{sacha_knowledge_2014}.

\vspace{-1mm}
\paragraph{Pros and Cons.} With this definition, insights are the culmination of a natural progression toward building the user's mental model of the data. It starts with extracting low-level data facts from a dataset, moves to group these facts into broader generalizations about the dataset, and finishes with the user formulating hypotheses to be tested in subsequent confirmatory analyses. As a result, this definition aligns well with the categorizations we analyzed in \autoref{sec:background:insight:categories}. This definition can also be considered an extension of the previous definition of insights as data facts. That being said, this definition also inherits the same limitations. Specifically, this definition completely omits the role of user experiences, domain knowledge, and social structures in forming insights. Further, it is still unclear how to construct the desired hierarchy from data facts to hypotheses without significant manual intervention~\cite{guo_case_2016}, requiring additional research in the future to formalize these structures.

\vspace{-1mm}
\subsubsection{Insights are Knowledge links}
\vspace{-1mm}

Finally, \revised{others argue that \textbf{insights can be seen as knowledge links.} For example,} Chang et al. \revised{say} that ``insight is considered to be more or less units of knowledge'' in visual analytics~\cite{chang_defining_2009}. \revised{This idea can also be extended to define} insights as \emph{links} \revised{connecting statistical and/or visual analysis \emph{findings} (e.g., \cite{srinivasan_augmenting_2019, kandogan2018towards})} with user \emph{knowledge}~\cite{saraiya_insight-based_2005,saraiya_insight-based_2006,north_toward_2006,shrinivasan_supporting_2008,rind_task_2016,gotz_interactive_2006,yi_understanding_2008,gotz_characterizing_2009,amar_knowledge_2005,amar_low-level_2005,smuc_score_2009,sacha_knowledge_2014,shrinivasan_connecting_2009,he_characterizing_2020,green_visual_2008,kandogan2018towards}, \revised{which can be synthesized from one or more user sessions~\cite{shrinivasan_supporting_2008,gotz_interactive_2006,saraiya_insight-based_2006,chang_defining_2009,gotz_characterizing_2009,smuc_score_2009,shrinivasan_connecting_2009,he_characterizing_2020,green_visual_2008}, a priori knowledge from outside the exploration process~\cite{saraiya_insight-based_2006,gotz_interactive_2006,yi_understanding_2008, pousman_casual_2007,amar_knowledge_2005,smuc_score_2009,sacha_knowledge_2014,shrinivasan_connecting_2009,he_characterizing_2020,green_visual_2008}, etc.}
\revised{Links can be implied, for example through observations made in qualitative insight studies}~\cite{saraiya_insight-based_2005,saraiya_insight-based_2006,smuc_score_2009}, or \revised{they can be digital objects made through annotation~\cite{gotz_interactive_2006,gotz_characterizing_2009,willett2011commentspace} and linking interactions~\cite{shrinivasan_supporting_2008,gotz_interactive_2006,gotz_characterizing_2009,shrinivasan_connecting_2009,he_characterizing_2020,dou_recovering_2009,willett2011commentspace} such as to connect system visualization state with the user's digitized notes.} \revised{Similarly}, insights can be \revised{made hierarchical and even composed together to form more complex insights}~\cite{saraiya_insight-based_2005,saraiya_insight-based_2006,north_toward_2006,pousman_casual_2007,gotz_characterizing_2009,smuc_score_2009,shrinivasan_connecting_2009,mathisen_insideinsights_2019,green_visual_2008}.

Moreover, Smuc et al. argue that insights can be more effectively analyzed through a direct analysis of how users' reported insights to build on one another and propose relational insight organizers (or RIOs) to organize and visualize the resulting insight graph~\cite{smuc_score_2009}.
RIOs share similarities with the structures proposed by Gotz et al.~\cite{gotz_interactive_2006}, where user knowledge is also captured as a graph, with high-level concepts and instantiations of these concepts stored as nodes within the graph, and relationships between instances stored as edges in the graph.
Similar graph-based structures have also been suggested by Shrinivasan and van Wijk~\cite{shrinivasan_supporting_2008}, Willett et al.~\cite{willett2011commentspace}, Mathisen et al.~\cite{mathisen_insideinsights_2019}, He et al.~\cite{he_characterizing_2020}, and Kandogan and Engelke~\cite{kandogan2018towards}.

\vspace{-1mm}
\paragraph{Pros and Cons.} We believe this definition is the most comprehensive conceptualization of insight since multiple units of knowledge and relationships between knowledge units can be represented using this definition. For example, data facts can be represented as units of analytic knowledge within a network graph, and domain knowledge can be represented as separate nodes within the graph for contextualizing these data facts.
That being said, most implementations of this definition keep insights at a relatively high level, for example, by only capturing domain knowledge as unstructured text in user-written notes or only capturing analytic knowledge as visualizations rather than the data characteristics that users \emph{interpreted} within these visualizations (e.g., specific correlations, differences in means, etc.). We still lack formal structures for reasoning about what we call the \emph{internal knowledge} that users extract from data and the \emph{external knowledge} users bring into an analysis session, hindering our ability to operationalize this definition in practice.

\vspace{-1mm}
\subsubsection{Integrating the Definitions}
\vspace{-1mm}

At face value, these definitions may appear distinct. However, a close look at the varying perspectives points to an overarching theme -- \textbf{an insight is a collection of knowledge}. We summarize the knowledge captured through insights in \autoref{fig:background:definitions:knowledge-units}. 
Although existing definitions vary in what they emphasize, e.g., prioritizing data facts versus domain knowledge, the components appear consistent across definitions, which we categorize as \emph{internal} and \emph{external knowledge}. For example, internal knowledge consistently includes data facts, generalizations, and hypotheses. External knowledge consistently includes domain expertise and personal experiences. Awareness of these components enables users of existing theory to navigate the varied definitions of insight; for example, identifying a definition emphasizing domain knowledge to motivate the design of a knowledge management tool~\cite{gotz_interactive_2006}.

\vspace{-1mm}
\subsection{Scoping Insights.}
\label{sec:background:insight:scope}
\vspace{-1mm}

Although there are many ways to express knowledge gained, gaining this knowledge generally occurs within a certain visual analysis scope~\cite{gotz_characterizing_2009,gathani2022grammar}. Further, \revised{scoping} insights appears to be tightly bound with \revised{defining} tasks~\cite{gathani2022grammar,brehmer_multi-level_2013} or objectives~\cite{rind_task_2016,lam_bridging_2018} in visualization research. Here, we explore how insights are scoped in the literature.

Many task models have been developed to categorize the scope of insights that analysts may be looking for.
These models often take the form of \emph{taxonomies} and \emph{typologies}~\cite{gathani2022grammar}, where tasks observed in the field or lab studies are generalized into abstract classes, such as ``Find Anomalies''~\cite{amar_low-level_2005}, ``Search/Comparison''~\cite{kang_examining_2012} or ``characterizing data distributions and relationships''~\cite{battle_characterizing_2019}.
Specific to insights, a number of taxonomies target insight generation \emph{processes} to understand whether task patterns may predict insight scope, rigor, and complexity~\cite{yi_understanding_2008,guo_case_2016,battle_characterizing_2019}.
Task models may also take the form of \emph{frameworks}, where the scope and structure of observed tasks, and relationships between these tasks, are abstracted into general-purpose hierarchies. Examples include the framework of tasks, sub-tasks, actions, and events proposed by Gotz and Zhou~\cite{gotz_characterizing_2009}, and the goals to tasks framework proposed by Lam et al.~\cite{lam_bridging_2018}.
We observe that these models predict the scope of insights by culling the set of relevant data facts (taxonomies) or narrowing the range of relevant data for applying these data facts (frameworks).
Further, these models seem to suggest an \emph{upper bound} on the depth and breadth of corresponding insights, where insights are unlikely to cover more data or facts than are predicted by these models. That being said, these models represent a range of possibilities. They are not meant to predict the \emph{exact} insights an analyst may uncover as they analyze a dataset.

An analyst's interest in pursuing certain tasks can also be defined \revised{by} the kinds of insights they \emph{expect} to uncover.
This observation stems from \revised{how} a user's analysis strategy is likely informed by an initial goal or ``hunch'' regarding the target dataset~\cite{lam_bridging_2018,zgraggen_investigating_2018,battle_characterizing_2019}, even \revised{if} vaguely at first~\cite{battle_characterizing_2019}.
For example, Bertin defines tasks according to the structure of the underlying data and the information the user seeks to learn from  this data~\cite{bertin1983semiology}. Andrienko and Andrienko extend Bertin's ideas to define tasks as declarative functions over data relations comprised of \emph{targets}, i.e., data attributes of interest, and \emph{constraints}, i.e., query predicates over these attributes~\cite{andrienko2006exploratory}. We note that Andrienko and Andrienko and Bertin's proposals overlap significantly with declarative definitions of task in database research, notably \emph{relational calculus},
a component of the \emph{relational model} that also defines tasks (or queries) as declarative functions over data relations~\cite{codd1970relational}.
That being said, existing declarative definitions of task are limited to scoping the user's \emph{expectations} and fail to encapsulate the insights that the user found, which are particularly interesting in insight discovery work.

Thus, although existing task models are useful aids for inferring insight scope, they alone are insufficient for fully defining insights and must be paired with alternative theories accordingly.

%% file: sections/discussion.tex
\section{Discussion}
\label{sec:discussion}
\vspace{-1mm}

In this section, we highlight promising directions for future research.

\vspace{-1mm}
\subsection{Building a ``Grammar of Insights''}
\vspace{-1mm}

The structural consistencies we observed across insight definitions draw close parallels to the role of grammars in visualization languages. For example, visualizations have been categorized in the same way that insights have~\cite{battle2018beagle,borkin2013what}. However, by favoring ease of use, visualization taxonomies also sacrifice the ability to express a diverse range of visualizations. Instead, one could identify the core building blocks behind them and construct grammars for expressing these building blocks. This is the core idea behind the Grammar of Graphics~\cite{wilkinson2012grammar}, which has led to wildly successful visualization grammars such as Vega-Lite~\cite{satyanarayan2017vegalite} and ggplot2~\cite{wickham2011ggplot2}.

We (and others~\cite{kandogan2018towards}) posit that the consistent structures observed across insight definitions suggests that one could also derive an equivalent ``grammar of insights,'' i.e., a unified formalism for expressing the core building blocks observed in our literature review. With a formalism, we can start to derive new grammars for expressing insights that maintain consistency with established definitions.
\revised{Declarative definitions of tasks~\cite{bertin1983semiology,andrienko2006exploratory,codd1970relational} and analytic knowledge~\cite{kandogan2018towards} provide promising starting points for new grammars, as do existing knowledge annotation systems (e.g., \cite{gotz_interactive_2006,shrinivasan_connecting_2009,willett2011commentspace}). However, we still lack generalizeable, programmable primitives for defining insights with equivalent precision to visualization grammars. As a first step towards filling this gap, we are developing a formalism for specifying insights and a corresponding grammar called Pyxis that can be used in visualization systems~\cite{battle2022programmatic}. Our ongoing work builds on this survey by specifying formal structures for the concepts we observe in the literature, such as internal knowledge (data facts, generalizations, etc.), external knowledge (domain knowledge, personal experiences, etc.), and links forged between the two to form insights (see \autoref{fig:background:definitions:knowledge-units}).}

\vspace{-1mm}
\subsection {Insights, Objectives and Tasks}
\vspace{-1mm}

An emergent theme from this work is that insights, objectives, and tasks are intertwined.  
For example, the types of insights gained during a task are likely influenced by the user's current objective, such as confirming an established hypothesis versus searching for interesting patterns in the data~\cite{battle_characterizing_2019}. Similarly, when a user pivots to a new task, i.e., changes their analysis objective, their recent insights likely influenced that pivot. The literature also seems to suggest that generalizable theory models should capture the various facets of user tasks (insights and objectives) and their \emph{interconnected nature}. For example, Andrienko and Andrienko model these connections by defining tasks in two parts~\cite{andrienko2006exploratory}: the target information sought during the task and the constraints the target must fulfill. Brehmer and Munzner expand on this principle through a multi-level typology that connects why the user performs a task, how they execute the related methods, and what the task’s inputs and outputs are~\cite{brehmer_multi-level_2013}. By connecting the inputs that guide the task (\emph{objectives}) with the outputs produced from the task  (\emph{insights}), we posit that theoretical task models could provide a holistic structure that enables researchers to analyze how tasks evolve and induce particular insights over time.

%% file: sections/conclusion.tex
\section{Conclusion}
\label{sec:conclusion}
\vspace{-1mm}

The paper presents a literature review charting the landscape of existing definitions of insight. We dissect the key components of these definitions, evaluate their pros and cons, and discuss their applicability to various scenarios in insight-based research. Based on our review, we identify two opportunities to extend existing theory: developing grammars of insight and connecting insight with theoretical models of visual analysis tasks and objectives.